\documentclass[aip,jcp,floatfix,reprint,showpacs,longbibliography,superscriptaddress]{revtex4-2}
\usepackage{amsmath}
\usepackage{url}
\usepackage{graphicx}
\usepackage{dcolumn}
\usepackage{scrextend}
\usepackage{hyperref}
\usepackage{bm}
\usepackage{amssymb}
\usepackage{rotating}
\usepackage[abs]{overpic}
\usepackage{xcolor}
\usepackage{tabularx}
\usepackage[labelsep=period]{caption}
\usepackage{floatrow}
\usepackage{subfig} 
\usepackage[justification=RaggedRight]{caption}
\usepackage{hyperref}
\usepackage{verbatim}
\hypersetup{colorlinks = true, citebordercolor={blue}, linkcolor={blue}, citecolor={blue}, urlcolor={blue}}
\begin{document}

\title{Beginnings of Exciton Condensation in Coronene Analog of Graphene Double Layer}

\author{LeeAnn M. Sager, Anna O. Schouten, and David A. Mazziotti}

\email{damazz@uchicago.edu}

\affiliation{Department of Chemistry and The James Franck Institute, The University of Chicago, Chicago, IL 60637 USA}%

\date{Submitted January 7, 2022\textcolor{black}{; Revised March 21, 2022}}


\begin{abstract}
Exciton condensation, a Bose-Einstein condensation of excitons into a
single quantum state, has recently been achieved in low-dimensional
materials including twin layers of graphene and van der Waals
heterostructures. Here we examine computationally the beginnings of
exciton condensation in a double layer comprised of coronene, a
seven-benzene-ring patch of graphene.  As a function of interlayer
separation, we compute the exciton population in a single coherent
quantum state, showing that the population peaks around 1.8 at distances
near 2 \AA.  Visualization reveals interlayer excitons at the separation distance
of the condensate.    We determine the exciton population as a function
of the twist angle between the two coronene layers to reveal the magic
angles at which the condensation peaks.  As with previous recent
calculations showing some exciton condensation in hexacene double layers
and benzene stacks, the present two-electron reduced-density-matrix
calculations with coronene provide computational evidence for the
ability to realize exciton condensation in molecular-scale analogs of
extended systems like the graphene double layer.

\end{abstract}

\maketitle

\section{Introduction}

Exciton condensation---a Bose-Einstein condensation of particle-hole pairs into a single quantum state---has generated considerable experimental and theoretical interest \cite{KSE2004,TSH2004,Fil_Shevchenko_Rev,Kogar2017,LWT2017,varsano_2017,fuhrer_hamilton_2016,Safaei2018,Sager2020,sager_53,Schouten2021a,SEP2000,KSE2002,KSE2004,TSH2004,NFE2012,LWT2017,Lee2016,Li2016,Li2017,tartakovskii_2019,GKY_2020} due to the resultant superfluidity \cite{london_1938,tisza_1947,DMA1995,AEM1995} of the constituent excitons (particle-hole pairs) allowing for the dissipationless transport of energy \cite{Keldysh2017, Fil2018}, which presents the possibility for uniquely energy efficient materials. Further, the greater binding energy and lesser mass of excitonic quasiparticles relative to particle-particle Cooper pairs indicates that exciton condensation should occur at higher temperatures \cite{FH2016} relative to the temperatures at which traditional superconductivity---i.e., the condensation of particle-particle pairs into a single quantum state \cite{BCS1957,Blatt_SC,Anderson_2013,Drozdov_250}---occurs.

Exciton condensates, nonetheless, have proven difficult to experimentally observe as excitons often have too short of a lifetime to allow for the simple formation of an exciton condensate; however, recent literature has established bilayer systems as being capable of demonstrating exciton condensation \cite{Min2008,PNH2013,Fogler2014,Li2016,Li2017,Ma2021,Hu_2020,Pikulin_2014,Debnath_2017,Burg_2018,Wang2019,Shimazaki_2020,Rickhaus_2021} likely due to the spatial separation of electrons and holes increasing excitonic lifetimes and causing them to act like oriented electric dipoles whose repulsive interactions prevent the formation of biexcitons and other competing exciton complexes such as electron-hole plasmas \cite{Eisenstein2004,Wang2019}.  Specifically, van der Waal heterostructures \cite{Sigl2020,Wang2019,Fogler2014, Kogar2017} as well as graphene bilayers \cite{Li2017,Liu2017, Min2008} demonstrate promise in the search for higher-temperature exciton condensate phases, with the tuneability of electronic states afforded by twisting graphene layers relative to each other being particularly of interest in recent literature \cite{Hu_2020,Rickhaus_2021}.

Small, molecularly-scaled systems have also been revealed to support exciton condensation via theoretical explorations utilizing a signature of such condensation found in the modified particle-hole reduced density matrix (RDM) \cite{Safaei2018,Sager2020,sager_53,Schouten2021a}. These molecular systems are able to be treated using theoretical approaches at lower computational costs and can be used as an analog for similar larger-scaled systems; moreover, molecular-scaled exciton condensation in and of itself may have potential applications in the design of more energy-efficient molecular-structures and devices.  As such, a coronene bilayer system \cite{Sancho_2014,Uehara_2021}---where each coronene layer is a seven-benzene-ring patch of graphene---is an ideal candidate for theoretical study of molecularly-scaled condensation phenomena. Exciton condensation in extended graphene bilayers indicate the likelihood that, similarly, coronene bilayers demonstrate correlation consistent with exciton condensation.  Conclusions drawn from such a study may prove useful in understanding the mechanism by which exciton condensation occurs in benzene-ring and graphene bilayers in general.

In this paper, we computationally examine the beginnings of exciton condensation in a double layer composed of coronene.  Utilizing variational 2-RDM theory \cite{Mazziotti2012, M2007, Mazziotti2011, Mazziotti2004, Mazziotti2016, Mazziotti_2020, nakata_nakatsuji_ehara_fukuda_nakata_fujisawa_2001, Mazziotti_2002, GM2008, Anthony_2016, Nik_2020, Nik_ACS_2020}, we explore the largest eigenvalue ($\lambda_G$) of the modified particle-hole reduced density matrix ($\tilde{G}$)---which corresponds to the largest population of excitons in a single particle-hole quantum state---for various coronene-bilayer geometries, such that an eigenvalue above the Pauli-like limit of one indicates exciton condensation as more than one exciton is occupying a single state and a larger eigenvalue indicates a higher degree of exciton condensate character.  We compare the maximal exciton populations ($\lambda_G$) as a function of distance between the layers of coronene and note that, near 2 \AA, the population peaks at around 1.8 with interlayer excitons being noted via our visualization technique at this distance.  Additionally, exciton populations as a function of twist angle between the two layers are computed in an effort to reveal any ``magic angles''.  Overall, this molecularly-scaled exploration of coronene bilayers provides computational evidence of the beginnings of exciton condensation in molecularly-scaled systems that is related to the condensation found in extended systems like graphene bilayers.

\section{Theory}

Condensation phenomena occur when bosons---or quasibosons---aggregate into a single, low-energy quantum ground state when adequately cooled \cite{bose_einstein_1924,einstein_1924}, which results in the emergence of superfluid properties \cite{london_1938,tisza_1947}.  For traditional bosons, a computational signature of so-called Bose-Einstein condensation occurs when the largest eigenvalue of the one-boson reduced density matrix (RDM)---expressed as
\begin{equation}
^{1} D^i_j = \langle \Psi |{ \hat b}^{\dagger}_i {\hat b}_j | \Psi \rangle
\label{eq:D1}
\end{equation}
where $|\Psi\rangle$ is an $N$-boson wavefunction and $\hat{b}^\dagger_i$ and $\hat{b}_i$ are bosonic creation and annihilation operators for orbital $i$, respectively---exceeds one \cite{Penrose_BEC}.  As the eigenvalues of the one-boson RDM correspond to the populations of one-boson orbitals, the largest eigenvalue corresponds to the maximum number of bosons occupying a single quantum state, i.e., the degree of condensation.

However, condensation in fermionic systems occurs via different mechanisms as multiple fermions cannot occupy a single orbital \cite{pauli_1940}.  In traditional superconductivity, superfluidity arises due to correlations within quasibosonic particle-particle (electron-electron, Cooper) pairs, causing the constituent Cooper pairs to flow without friction \cite{BCS1957,Anderson_2013}.  The signature of particle-particle condensation is the largest eigenvalue of the particle-particle RDM (2-RDM) \cite{Y1962,S1965} given by
\begin{equation}
^{2} D_{k,l}^{i,j} = \langle \Psi | {\hat a}^{\dagger}_i {\hat a}^{\dagger}_j {\hat a}_l {\hat a}_k  | \Psi \rangle
\label{eq:D2}
\end{equation}
where $|\Psi\rangle$ is an $N$-fermion wavefunction, where each number demonstrates both the spatial and spin components of the fermion, the indices $i,j,k,l$ correspond to one-fermion orbitals in a finite basis set of rank $r$, and $\hat{a}^\dagger$ and $\hat{a}$ depict fermionic creation and annihilation operators, respectively.  The largest eigenvalue of the 2-RDM corresponds to the largest population of a single particle-particle quantum state (called a geminal \cite{Y1962,C1963,S1965,RM2015,Srev_1999,shull_1959}), i.e., the degree of particle-particle condensation.

Similarly, exciton condensation results from particle-hole pairs (excitons) condensing into a single quantum state \cite{Fil_Shevchenko_Rev,keldysh_2017}.  The signature of exciton condensation---denoted as $\lambda_G$---is a large eigenvalue ($\lambda_G > 1$) of a modified version of the particle-hole reduced density matrix \cite{Safaei2018, GR1969, Kohn1970}, with elements given by
\begin{multline}
{}^{2}\Tilde{G}^{i,j}_{k,l}={}^{2}G^{i,j}_{k,l}-{}^{1}D^i_j{}^{1}D^l_k \\=\langle \Psi | {\hat a}^{\dagger}_i {\hat a}_j {\hat a}^{\dagger}_l{\hat a}_k  | \Psi \rangle-\langle\Psi|\hat{a}^\dagger_i\hat{a}_j|\Psi\rangle\langle\Psi|\hat{a}^\dagger_l\hat{a}_k|\Psi\rangle
\label{eq:modG2}
\end{multline}
where ${}^{1}D$ is the one-fermion reduced density matrix (1-RDM).  After modification---which removes an extraneous ground-state-to-ground-state transition---the largest eigenvalue of the particle-hole RDM corresponds to the number of particle-hole pairs (excitons) that occupy a single particle-hole quantum state and hence signifies presence and extent of exciton condensation.

\section{Results}

Extended graphene bilayer systems have been identified as a major candidate for the creation of macroscopically-scaled exciton condensates \cite{Li2017,Liu2017, Min2008}. In  this  study,  we  extrapolate  this  framework to a molecularly-sized system  and  use  bilayers of coronene---where each layer is composed of seven, joined benzene rings---in order to probe a molecularly-scaled system whose similarity to graphene bilayers make it both a promising contender for a molecularly-scaled exciton condensate as well as an ideal analogue for exploring the correlation in layers of graphene using a system that can be directly explored by current theoretical techniques for strong electron correlation.  As such, we explore relative amounts of correlation in coronene bilayer systems as a function of both interlayer distance and twist angle.

\subsection{Exciton Population with Distance}

To gauge the relative extents of exciton condensation as a result of varying interlayer distances between each layer of coronene and thus probing a significant range of van der Waals interactions between each layer in an attempt to identify an ideal distance for maximal correlation, interlayer spaces are varied from 1.0 \AA \ to 2.5 \AA, and the signature of condensation---$\lambda_G$, i.e., the number of excitons condensed into a single particle-hole quantum state---is probed in the STO-6G basis using variational 2-RDM theory with a [24,24] active space.

\begin{figure}[tbh!]
    \centering
    \includegraphics[width=8cm]{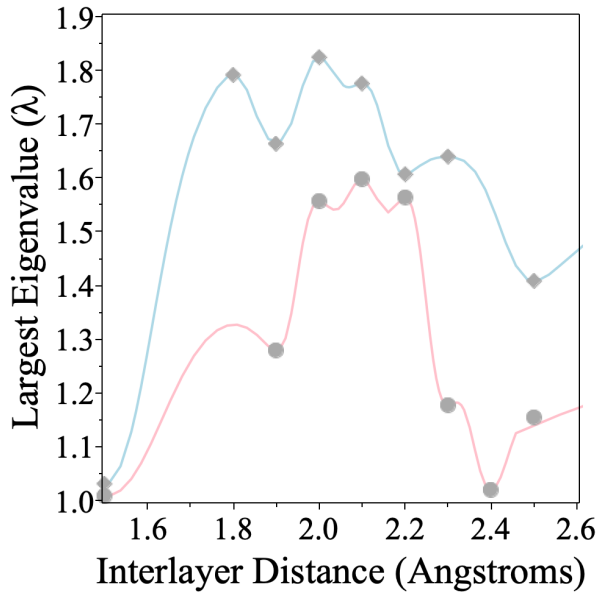}
    \caption{A scan over the exciton population in a single coherent quantum state (i.e., the largest eigenvalue of the modified particle-hole RDM) versus the distance between the two coronene layers for V2RDM-CASSCF calculations using a [24,24] active space (blue) and CI-CASSCF calculations using a [10,10] active space (pink).  A STO-6G basis is utilized for both calculations.}
    \label{fig:24_24_dist_scan_sto}
\end{figure}

As can be seen in Fig. \ref{fig:24_24_dist_scan_sto}---where the blue data indicates variational 2-RDM complete-active-space self-consistent-field (V2RDM-CASSCF) [24,24] calculations with [X,Y] denotes an active space of X electrons in Y orbitals and the pink data indicates configuration-interaction-based complete-active-space self-consistent-field (CI-CASSCF) [10,10] calculations---coronene bilayer systems demonstrate character of exciton condensation ($\lambda_G>1$) for a wide variety of interlayer distances with the maximal excitonic populations in the bilayer peaking at 1.824 at 2 \AA, although a relatively-wide plateau is noted in the range of 1.8-2.2 \AA, indicating that exciton condensation is relatively robust in that region of distances.  Distances in the neighborhood of 2.0 \AA \ are hence ideal for the study of exciton condensate phases in bilayer systems composed of coronene, at least for twist angles around 0 degrees.

\begin{figure*}[tb!]
    \centering
     \subfloat[1.5 \AA, $\lambda_G=1.031$]{\label{fig:vis_a}\includegraphics[width=5.4cm]{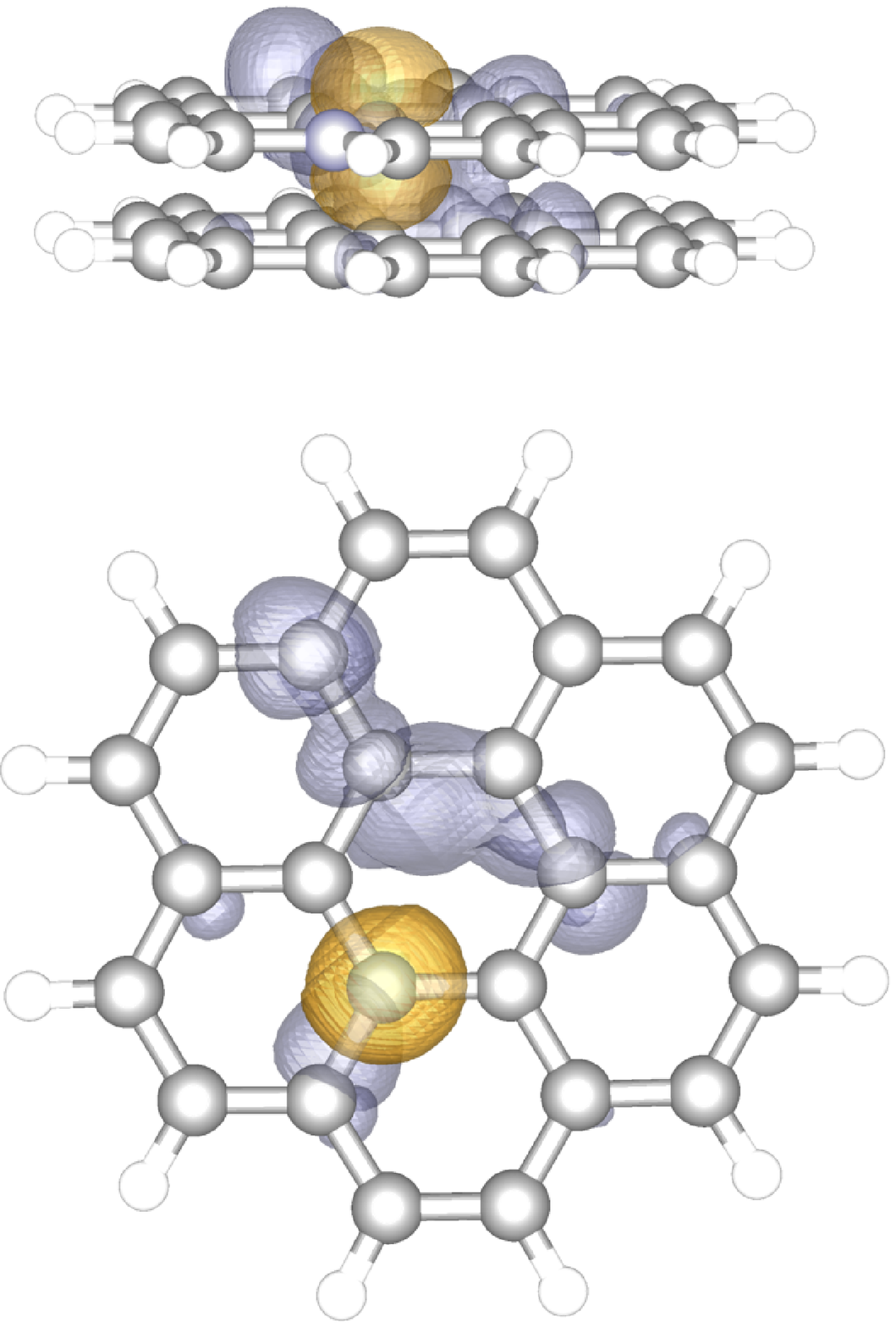}}
     \hspace{0.1cm}
     \subfloat[2.0 \AA, $\lambda_G=1.824$]{\label{fig:vis_b}\includegraphics[width=5.4cm]{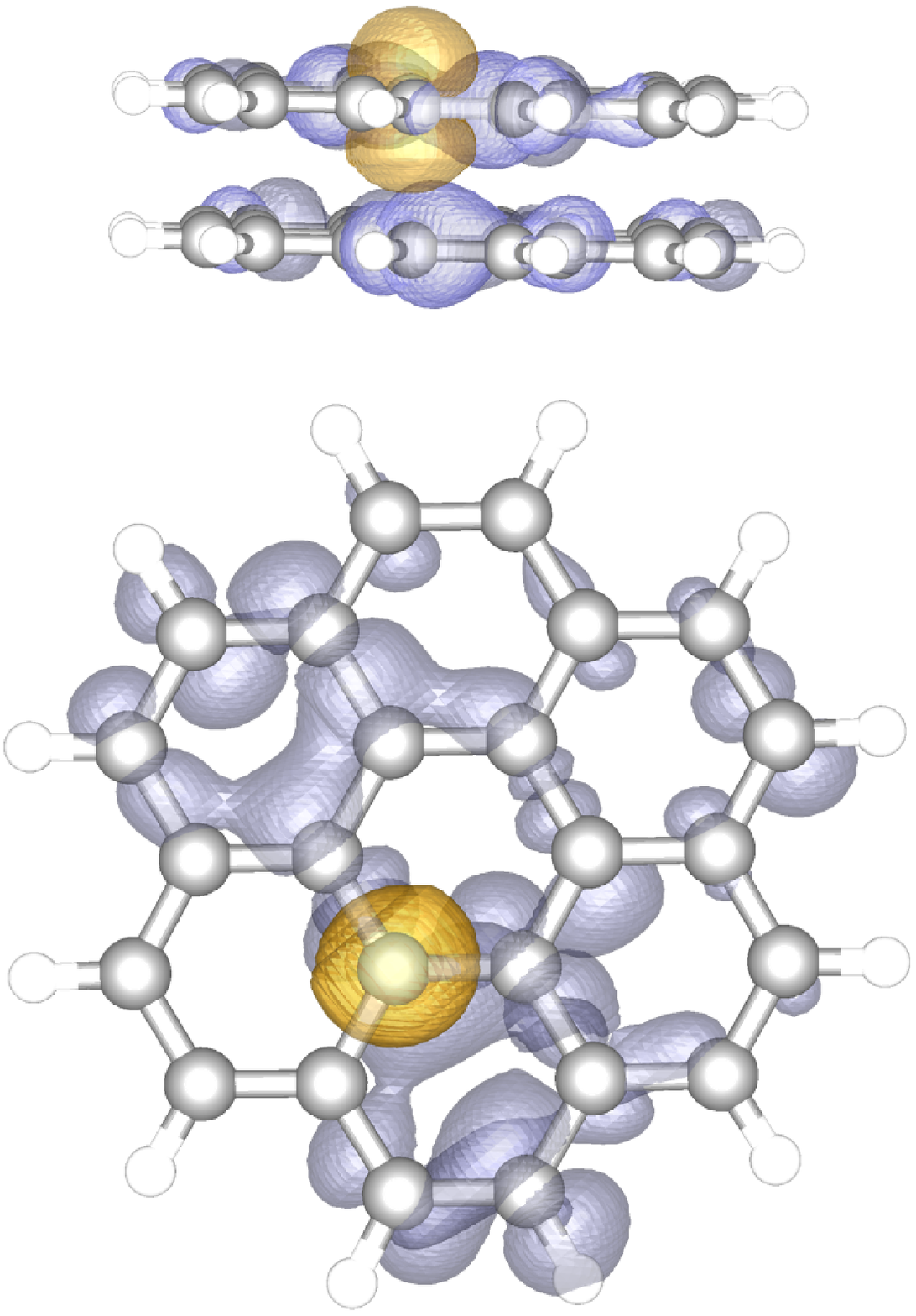}}
     \hspace{0.1cm}
     \subfloat[2.5 \AA, $\lambda_G=1.408$]{\label{fig:vis_c}\includegraphics[width=5.4cm]{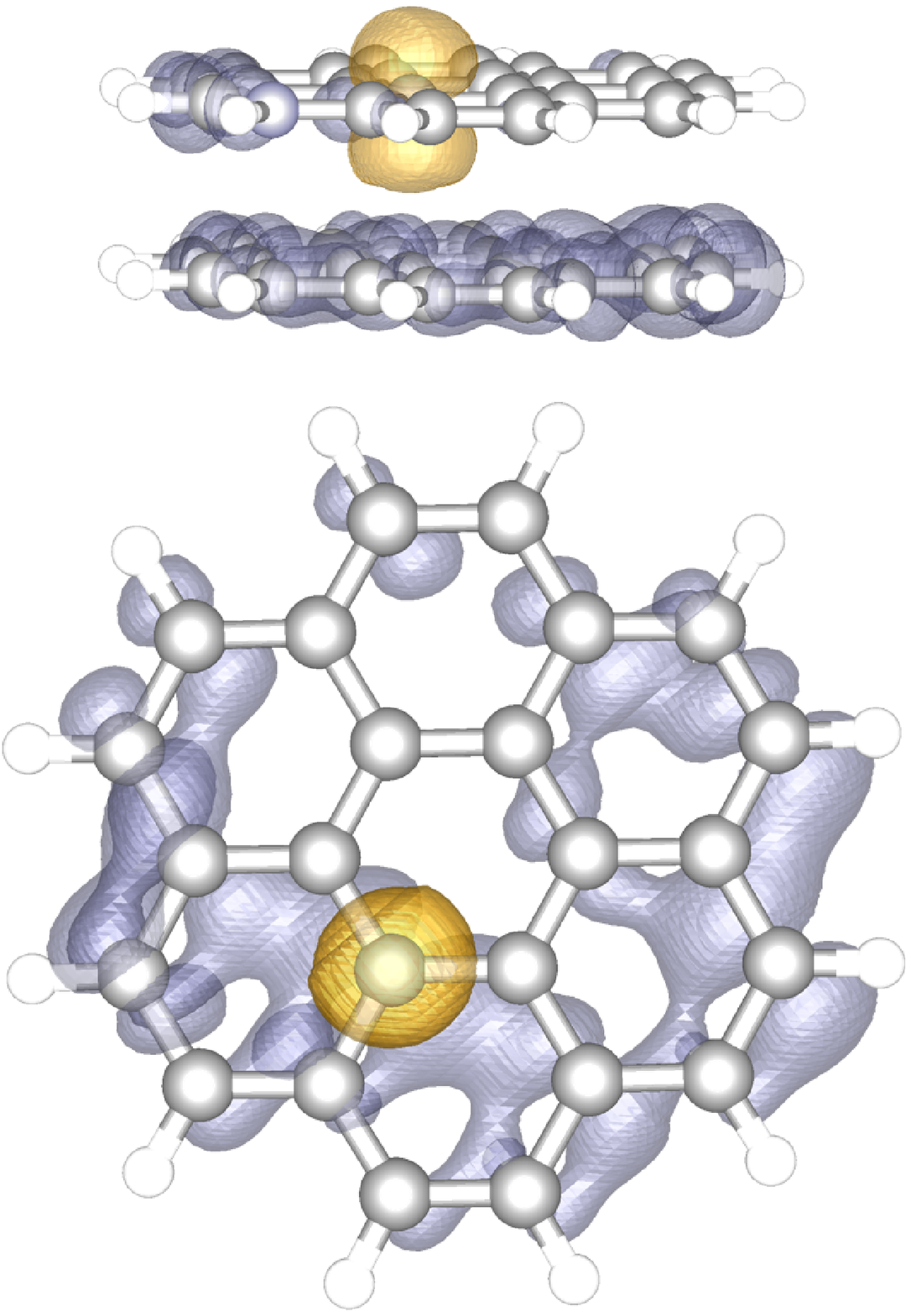}}
    \caption{Visualizations of the non-rotated coronene bilayer systems for (a) 1.5 \AA, (b) 2.0 \AA, and (c) 2.5 \AA \ where the gray-violet represents the probabilistic location of the hole in the particle-hole wavefunction associated with the large eigenvalue for a particle position in a fixed atomic orbital (gold).  
    Variational 2-RDM calculations with a [24,24] active space and STO-6G basis set are utilized for each visualization.}
    \label{fig:orb_visualizations}
\end{figure*}

Figure~\ref{fig:orb_visualizations} allows for the visualization and comparison of coronene bilayer systems with differing degrees of excitonic condensation.  For the particle-hole wavefunction associated with the large eigenvalue, we visualize the probability distribution of the hole (gray-violet) for a particle location in a $2p_z$ orbital of one of the symmetrically-equivalent carbon atoms in the interior benzene ring (gold) using the methodology described in \ref{sec:vis_tech}
for each geometry.  The density cut-off for the probabilistic location of the hole differs between all three visualizations, 
so the magnitudes of the densities can not be directly compared between computations; however, the general trends in hole density locations can be established.  For the 2.0 \AA \ calculation, which shows the maximal excitonic character of the set, the excitonic hole is highly delocalized between both layers, demonstrating a highly correlated interlayer exciton; for the 2.5 \AA \ calculation, an interlayer exciton is still observed, however, the degree of delocalization is highly decreased---with the majority of the hole population being focused on a single layer---, consistent with a lower degree of correlation and a lower signature of condensation; finally, for the 1.0 \AA \ calculation, which does not demonstrate any exciton condensation, the hole's probabilistic location is highly localized with the majority of the population in the same layer as the particle.  As such, the delocalization and the interlayer location of the hole seem to be strong indicators of high degrees of exciton condensation and indicate that both factors may be necessary for a condensate to form.

Note that, while the signature of exciton condensation is depressed in the [10,10] CI-CASSCF calculations relative to the [24,24] V2RDM-CASSCF calculations---which is expected as the higher active spaces allow for higher degrees of correlation, which can lead to higher signatures of condensation---the overall trends between the two sets of data are consistent, especially in the region of maximal condensation.  As the results are consistent and as the [10,10] calculations are less computationally-expensive, [10,10] CI-CASSCF calculations are used throughout the exploration of the effect of twist angles on the presence and extent of exciton condensation.

\subsection{Exciton Population with Twist Angle}

To obtain a more complete understanding of the conditions under which exciton condensation occurs, we explore the effect of rotations between the two layers of coronene on the excitonic population in a single quantum state (i.e., $\lambda_G$).

\begin{figure*}[tb!]
    \centering\subfloat[2.0 \AA, Large Angle Scan]{\label{fig:angle_a}\includegraphics[width=5.5cm]{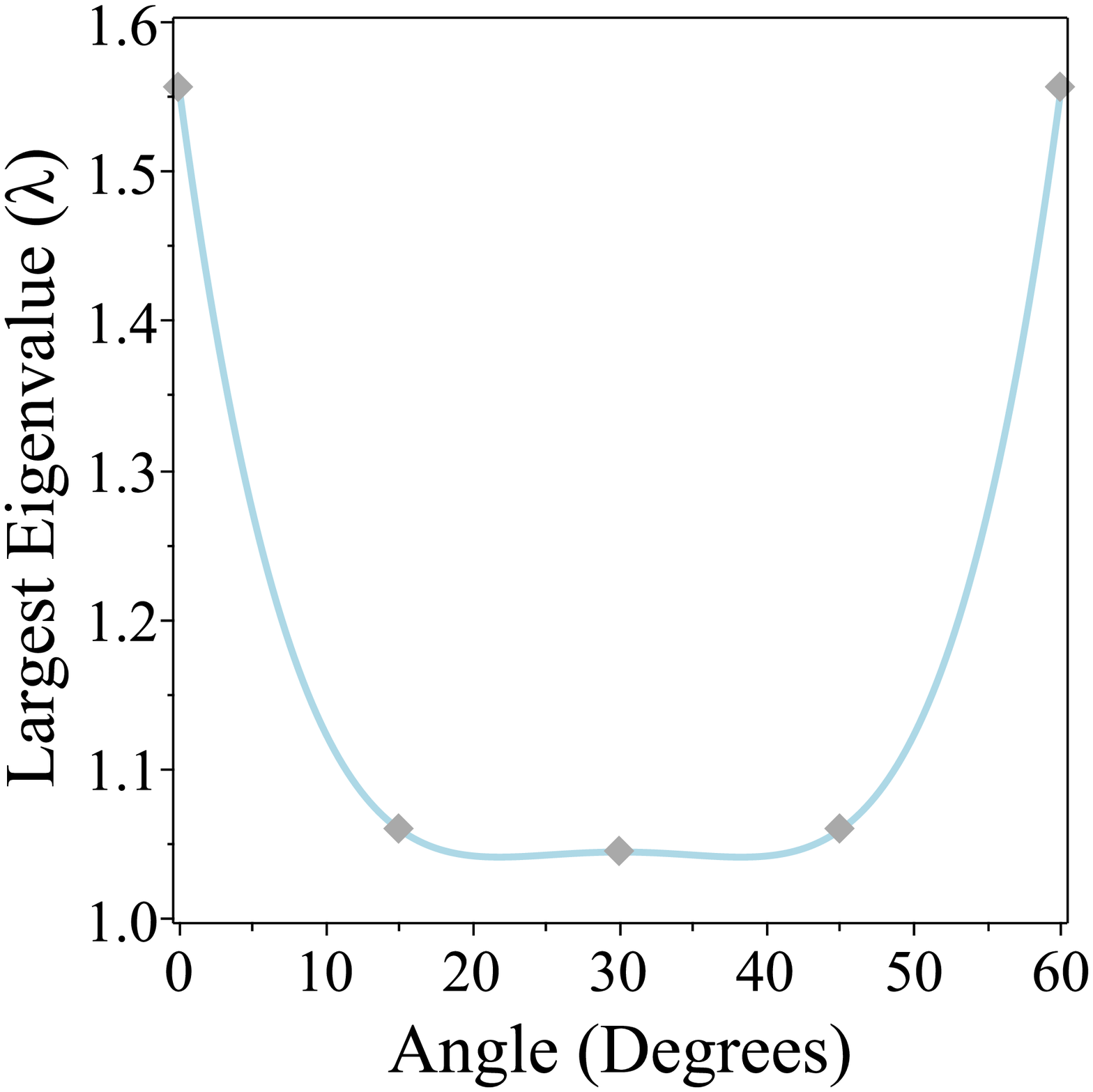}}
     \hspace{0.1cm}
     \subfloat[2.0 \AA, Small Angle Scan]{\label{fig:angle_b}\includegraphics[width=5.5cm]{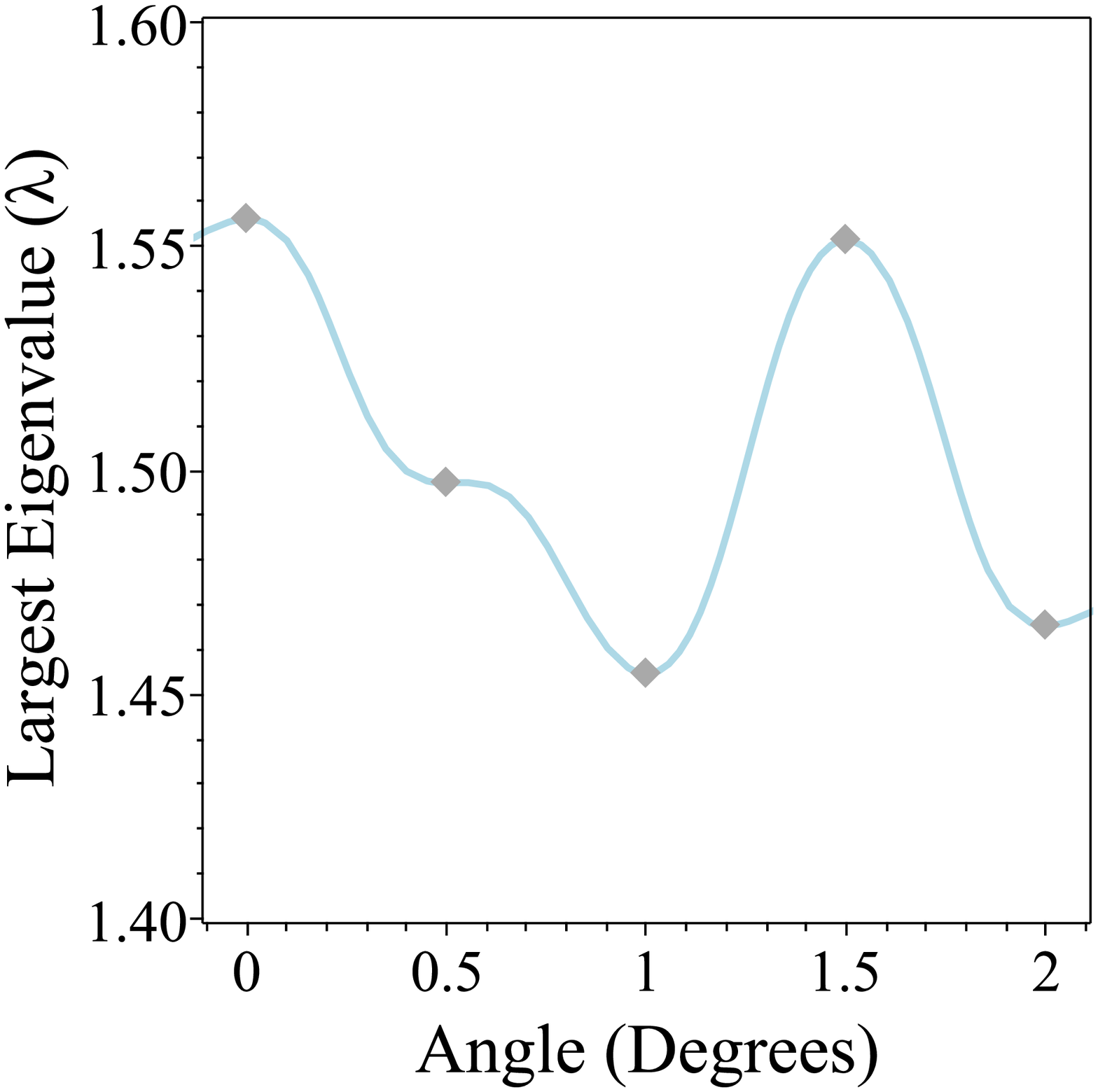}}
     \hspace{0.1cm}
     \subfloat[0(blue)/15(pink)/30(green) Degrees,  Distance Scan]{\label{fig:angle_c}\includegraphics[width=5.5cm]{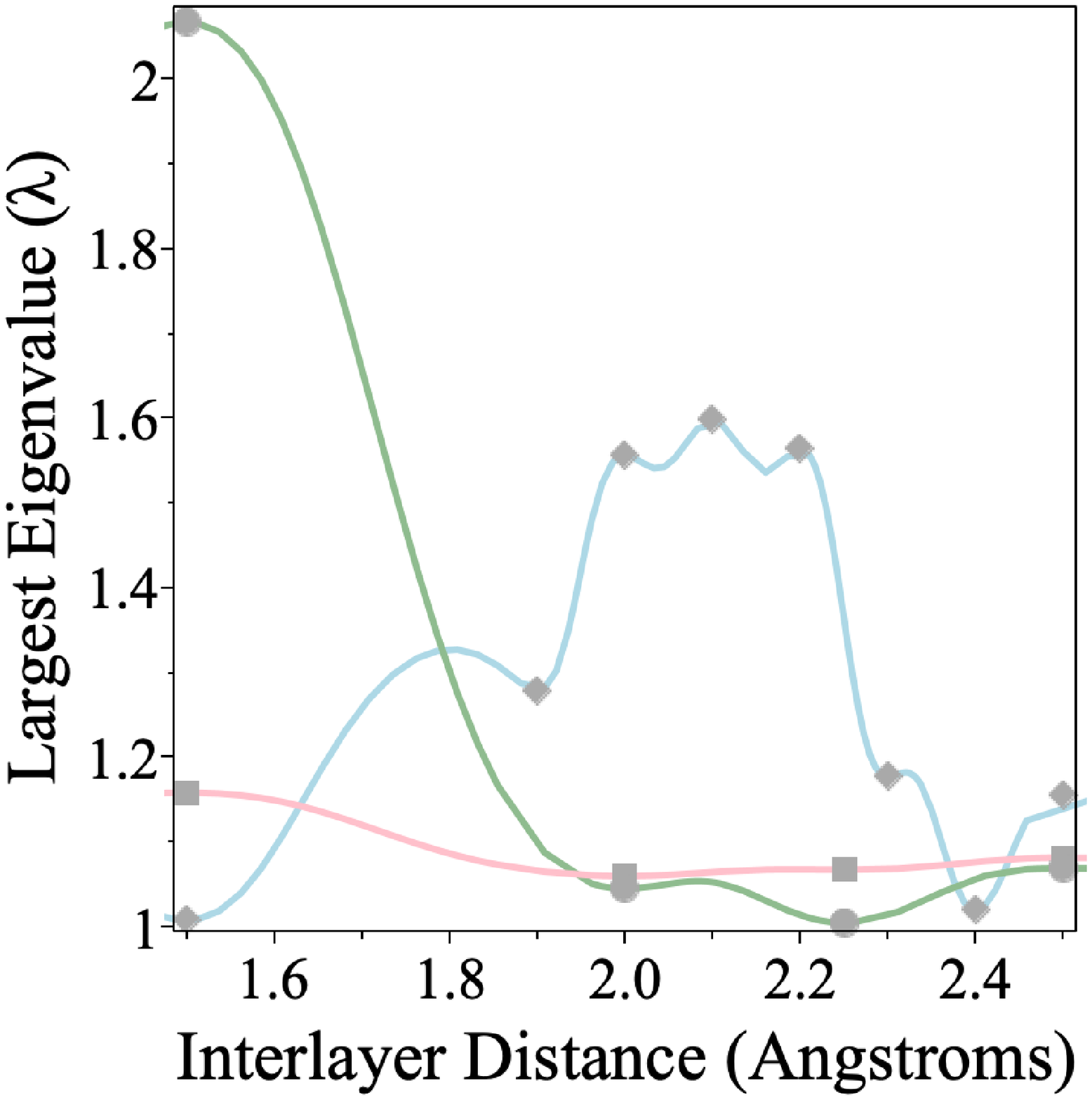}}
    \caption{A scan over the exciton population in a single coherent quantum state (i.e., the largest eigenvalue of the modified particle-hole RDM) versus (a)small or (b)large angle variations are shown in the left-most and middle figure, and a scan over exciton population versus interlayer distance for twist angles of 0 (blue), 15 (pink), and 30 (degrees) is shown in the right-most figure. Activespace SCF calculations with a [10,10] active space with a STO-6G basis are utilized for each plot.}
    \label{fig:angle_scan}
\end{figure*}

As can be seen from Fig. \ref{fig:angle_a}---which scans the exciton population as a function of angles from 0 to 60 degrees, the full range of rotation before an identical configuration is obtained, for coronene bilayer systems with an interlayer distance of 2.0 \AA---maximal condensation character is noted in the range of no offset.  However, as shown in Fig. \ref{fig:angle_b}, the large degree of condensation is relatively stable in the region of small angles, particularly of interest in magic-angle graphene studies [at around 1.1 degrees, \cite{Cao_2018}], with the largest degree of condensation occurring at 0 degrees but with all angles scanned---from 0 degrees to 2 degrees by 0.5 degrees---the maximal exciton population remains above 1.45.

Additionally, in order to determine whether the optimal interlayer distance is consistent between different twist angles, Fig. \ref{fig:vis_c} shows a scan of the degree of condensation versus the distance between the two coronene layers for twist angles of 0 (blue), 15 (pink), and 30 (green) degrees. Interestingly, the optimal interlayer distance for both the 15 and 30 degree twist angles is decreased from 2.0 \AA \ to 1.5 \AA, with the 15 degree maximum at 1.5 \AA \ being significantly decreased from that for the unrotated maximum at 2.0 \AA \ and the 30 degree maximum at 1.5 \AA \ being significantly higher---showing a maximal exciton population above two, indicating that more than two excitons are occupying a single quantum state, even using the [10,10] active space with fewer degrees of correlation relative to the [24,24] active space previously used to scan degree of condensation versus interlayer distance for the unrotated bilayer system in the analysis in the preceding section.

\section{Discussion and Conclusions}

In this study, we theoretically probe the presence and extent of exciton condensation---via the use of a quantum signature measuring the exciton population of a single particle-hole quantum state---for a variety of coronene bilayers.  In these coronene bilayer systems---which are molecularly-scaled analogues of extended graphene bilayer systems---, we optimize the excitonic character versus the distance between the bilayers and find an excitonic populations of around 1.8 for interlayer distances around 2.0 \AA \ when the coronene layers have a twist angle of zero degrees; this signature of condensation is seen to be relatively robust in the region of 1.8-2.3 \AA, which while shorter than experimental bilayer distances of around 3.0 \AA \ \cite{Uehara_2021}, may be attainable using either an appropriate linker or high pressure.

Further, by exploring the effect of the angle of rotation between the two coronene layers (i.e., the twist angle), we discover that for distances around 2.0 \AA, the optimal twist angles between the layers are those corresponding to completely-aligned layers (i.e., 0, 60, 120, etc. degrees), although this maximal condensate character is rather robust for small angles around those explored in magic-angle graphene studies \cite{Cao_2018}.  Moreover, by investigating the relationship between interlayer distance and excitonic populations for different twist angles, we note a large dependence on the degree of rotation on the signature of condensation.  Specifically, we find the overall highest degree of exciton condensation---with an excitonic population above two in a single quantum state---for a coronene bilayer geometry corresponding to a 30 degree twist angle and a 1.5 \AA \ interlayer distance, which is slightly shorter than a carbon-carbon single bond and hence may not be an experimentally-feasible distance.  As such, an experimental exploration of molecular-scaled exciton condensation in coronene bilayer systems should likely focus on untwisted bilayer geometries in the range of 2.0 \AA.

Interestingly, our visualization technique in which an exciton---corresponding to the largest degree of condensation---is visualized by plotting a the hole's probabilistic location for a specified particle location, indicates that interlayer excitons may be required in order for the coronene bilayer system to demonstrate exciton condensation.  For visualizations with the same specified particle orbital (the $2p_z$ orbital on one of the symmetrically-equivalent carbon atoms in the interior benzene ring), geometries demonstrating character of exciton condensation have clear, delocalized, interlayer excitons.  Further, we note that the delocalization of the hole location increases with the increase in excitonic population.

\begin{figure*}
    \centering
    \includegraphics[width=13cm]{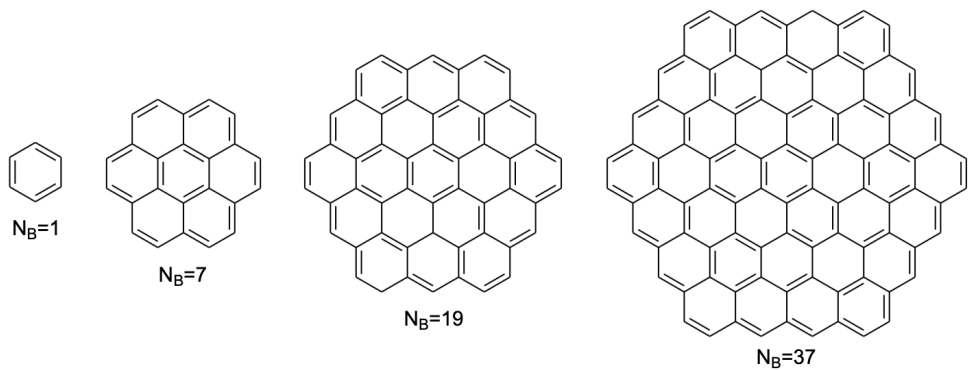}
    \caption{\color{black}The progression of molecular systems in which benzene is completely surrounded by (left to right) zero, one, two, and three complete rows of benzene are shown.  These molecules---which have one, seven, nineteen, and thirty-seven constituent benzenes, respectively---would be necessary to extrapolate exciton condensate behavior to the limit of an extended layer of graphene.\color{black}}
    \label{fig:series}
\end{figure*}

\color{black}
An interesting future direction may be the exploration of trends in exciton condensation with system size.  Such a study would likely be beneficial in extrapolating smaller-system exciton condensation results to predict behavior of extended system graphene.  In prior work, we have indeed explored the relationship between system size---i.e., number of benzene units---in both horizontal bilayer systems including pentacene and hexacene \cite{Safaei2018} as well as vertically in multilayer, molecular-scale van der Waals stacks composed of benzene subunits with the latter demonstrating an almost-linear increase of condensate character with an increase in the number of layers \cite{Schouten2021a}. In the case of coronene, however, such an extrapolation would prove difficult with current theoretical methodologies robust enough to capture the correlation phenomena inherent to condensation behavior as the natural progression of molecules---shown in Fig. \ref{fig:series}---rapidly become prohibitively large especially considering double-layer systems.

Another interesting direction would be determination of the temperature dependency of excitonic phenomena.  Exciton condensation in coronene is a ground state phenomenon in which multiple constituent particle-hole quasibosons condense into a single quantum state; as such, we would expect it to persist at finite and small temperatures up until some critical temperature at which thermal energy is sufficient to disrupt the condensation.  Determination of critical temperature can be explored theoretically, although the calculation of excited states would be required, which may be an interesting future avenue of exploration.
\color{black}

Overall, this study identifies a candidate for molecular-scale exciton condensation---namely, a bilayer of coronene subunits with twist angles near zero degrees and interlayer separations near 2 \AA---, which could have applications in applications involving molecularly-scaled electronic structures and devices.  Further, the clear signature of exciton condensation noted for the molecular-scale analogue of a graphene bilayer supports the idea that the interesting electronic phenomena in graphene bilayer systems could be occurring via an excitonic mechanism.  The understanding gained throughout this geometric analysis of coronene bilayers illuminates the relationships between twist angle, interlayer distance, and degree of exciton condensation, which increases our understanding of geometric considerations in the design of graphene bilayer-like exciton condensate materials.

\begin{acknowledgments}
D.A.M. gratefully acknowledges support from the U. S. National Science Foundation Grant No. CHE-1565638 and DGE-1746045 and the ACS Petroleum Research Fund Grant No. PRF No. 61644-ND6.
\end{acknowledgments}

\section*{Conflict of Interest}
The authors do not have a conflict of interest to report. \\

\section*{Data Availability Statement}
The data is available from the corresponding author upon reasonable request. \\

\appendix

\section{Computational Methods}

The particle-particle reduced density matrix (2-RDM) for the coronene bilayers is obtained directly from the molecular structure using a variational method \cite{ Mazziotti2012, Mazziotti2004, mazziotti_2016, nakata_nakatsuji_ehara_fukuda_nakata_fujisawa_2001, M2007, Mazziotti1998}. Additional constraints allowing the 2-RDM to represent $N$-particle wavefunctions---i.e., $N$-representability conditions---, require the the particle-particle, hole-hole, and particle-hole RDMs all to be positive semi-definite. The STO-6G basis set is used for all coronene bilayer calculations, and the active space utilized is specified throughout the document, with---unless otherwise noted---[24,24] variational 2-RDM CASSCF calculations being utilized for the scan over interlayer distances and [10,10] CI-CASSCF calculations being utilized for the scan over twist angles.

The 2-RDM obtained directly from the molecular structure is then utilized to construct the particle-hole RDM by the linear mapping given by:
\begin{equation}
 ^{2}G^{i,j}_{k,l} = {\delta^{j}_{l}}^{1}D^{i}_{k} + ^{2}D^{i,k}_{j,l}.
\end{equation}
The modified particle-hole RDM can then be obtained from the particle-hole RDM according to:
\begin{equation}
    {}^{2}\tilde{G}^{i,j}_{k,l}={}^{2}G^{i,j}_{k,l}-{}^{1}D^i_j{}^{1}D^l_k.
\end{equation}
The eigenvalues ($\lambda_{G,i}$) and eigenvectors ($\overrightarrow{v}_{G,i}$) of the modified particle-hole matrix are calculated using an eigenvalue optimization:
\begin{equation}
    \tilde{G}\overrightarrow{v}_{G,i} = \lambda_{G,i}\overrightarrow{v}_{G,i}
\end{equation}
where the largest eigenvalue of the modified particle-hole RDM is the signature of condensation that represents the largest exciton population in a single coherent quantum state.

\section{Visualization Technique}
\label{sec:vis_tech}

The ``exciton density'' visualization shows the probabilistic hole location as a function of a specific particle location.  This information is obtained via a matrix of atomic orbitals in terms of molecular orbitals, $M_\mathrm{{AO,MO}}$, calculated directly from a matrix of molecular orbitals in terms of atomic orbitals, $M_\mathrm{{MO,AO}}$, which is obtained as an output of the direct computation of the 2-RDM:
\begin{equation}
    M_\mathrm{{AO,MO}} = (M^{T}_\mathrm{{MO,AO}})^{-1}.
\end{equation}
A submatrix corresponding to the active orbitals is isolated from the overall matrix, and the eigenvector of the largest eigenvalue of the modified particle-hole RDM is reshaped as a matrix in the basis of the active orbital submatrix. The eigenvector matrix, denotes by $V_\mathrm{{max}}$, is then utilized to create
\begin{equation}
    (M^\mathrm{{active}}_\mathrm{{AO,MO}})(V_\mathrm{{max}})(M^\mathrm{{active}}_\mathrm{{AO,MO}})^{T}.
\end{equation}
which is a matrix representing electron atomic orbitals in terms of the corresponding probabilistic hole location, with the resultant coefficients contributing to other orbitals.


%

\end{document}